\documentclass[runningheads]{svmult}
\usepackage{graphicx}  

\begin{document}
\title*{Nonlinear Dynamics of Active Brownian Particles} 
\toctitle{Nonlinear Dynamics of Active Brownian Particles}
\titlerunning{Active Brownian Particles}
\author{Werner Ebeling\inst{1}}
\authorrunning{Werner Ebeling}
\institute{Institute of Physics, Humboldt--University Berlin, 
        Invalidenstr.110, D--10115 Berlin, Germany} 
\date{\today}
\maketitle

\begin{abstract}
We consider finite systems of interacting Brownian particles 
including active friction in the framework of nonlinear dynamics 
and statistical/stochastic theory. 
First we study the statistical properties for $1-d$ systems of masses 
connected by Toda springs which are imbedded into a heat bath. 
Including negative friction we find $N+1$ attractors of motion
including an attractor describing dissipative solitons.
Noise leads to transition between the deterministic attractors. 
In the case of two-dynamical motion of interacting particles 
angular momenta are generated and left/right rotations of pairs and swarms
are found.   
\end{abstract}


\section{Introduction}

The main purpose of this work is the study of the dynamics of active 
Brownian particles including interactions. The friction is modelled by a
velocity-dependent  function derived from a model of energy supply
\cite{ScEbTi98,EbScTi99,ErEbSS00,MaEbVe00,EbErDJ00}.
The interaction between the particles is modelled by Toda potentials 
(in $1-d$) or Morse potentials (in $2-d$).\\  
Since the classical work of Toda
\cite{Toda81} the development of the nonlinear dynamics and statistical
thermodynamics of of Toda systems has remained a central topic of research.
Toda was able to find exact solutions for a special $1-d$ system with a
exponential interaction. In particular Toda proved the existence of soliton
solutions and calculated the exact partition function. Since that solitons as
excitations of nonlinear chains of masses found a remarkable interest. Several
interesting results were obtained for the statistical thermodynamics of Toda
systems  \cite{KrSc75,BeSc81,TrZaPo86}. 
We consider here systems with active friction and noise by means of 
analytical tools and simulations. 
It was shown in previous work, that there
exists a special temperature regime around a transition temperature $T_{tr}$
between the phonon and the soliton regime, in another context called
the localization temperature $T_{loc}$, where the interaction of solitons
and phonons is strong and has a remarkable influence on several physical phenomena,
a special one being energy localization at special sites \cite{EbJe91} and another
one the excitation of a broadband coloured noise spectrum with an $1/f$ region
at low frequencies \cite{JeEb00}. Here our main interest is devoted to the
influence of negative friction on the properties of Brownian motion. 
Driving the system by negative friction to far from equilibrium states 
we find $N+1$ attractors of deterministic motion including an attractor
describing dissipative solitons. Noise leads to transition between the
deterministic attractors. In the case of two-dynamical motion of interacting
particles positive or negative angular momenta are generated with equal
probability. This leads to left/right rotations of pairs, clusters and 
swarms. We will show that the collective motion of large clusters
of driven Brownian particles reminds very much the typical modes
of parallel motions in swarms of living entities.     

\section{Equations of motion, friction and forces}

Let us consider a systems of $N$ point masses $m$ with the numbers
$1,2,...,i,...N$. We assume that the mass $m$ is connected to the next
neighbours at both sides by Toda forces,  
The distance between the mass $i$ and the mass $i+1$ is denoted by $R_i$, the
equilibrium distance is assumed to be $\sigma$, therefore the spring
elongation reads $r_i=R_i-\sigma$. In the following 
take $\sigma$ as the length unit. 

The dynamics of the system is given by the
following equation of motion for the elongations
\begin{equation}
\frac{d^2}{dt^2} r_i= \left[V'(r_{i+1}) - 2 V'(r_{i}) + V'(r_{i-1})\right]-
\gamma(v_i) v_i + {\cal F}_i (t)
\end{equation}
where 
${\cal F}_i (t)$ is a stochastic force with strength $D$ and a
$\delta$-correlated time dependence:
\begin{equation}
\label{stoch}
\langle {\cal F}_i (t) \rangle=0 \,;\,\,
\langle {\cal F}_i(t){\cal F}_j (t')\rangle=2D \,\delta(t-t') \delta_{ij}
\end{equation}

In the case of thermal equilibrium systems we have 
$\gamma(v)=\gamma(0)={\rm const.}$. In the general case where
the friction is velocity dependent we will assume that the friction is
monotonically increasing with the velocity and that 
the limit for large
velocities is a well defined constant:
\begin{equation}
\gamma(v) \rightarrow \gamma(0)={\rm const.}
\end{equation}
Our basic assumption is, that in this limit
the loss of energy resulting from friction, and the gain of energy
resulting from the stochastic force, are compensated in the average.
From this postulat follows the 
non-equilibrium fluct\-uation-dissipation theorem :
\begin{equation}
  \label{fluct-diss}
  D=k_{B}T\gamma_{0}/m
\end{equation}

Here $T$ is the temperature of the heat bath, $k_{B}$ is the Boltzmann
constant, and $D$ is the strength of the stochastic force.
For the case of a passive thermal heat bath $\gamma = \gamma_0$
eq.(\ref{fluct-diss}) agrees with the conventional Einstein
relation. The validity of a fluctuation-dissipation relation between
noise strength and damping strength is assumed in order to guarantee the
excistence of a stationary or thermal equilibrium, independent on the 
limit of the friction
parameter $\gamma_0 \geq 0$. We note that $\gamma_0 =0, T = 0$ corresponds to
the conservative case.

In the following we will study first the simplest case of passive friction,
i.e. we will assume that the friction
function is constant $\gamma(v)=\gamma_0 = const$. Then more complicated
friction functions will be studied. Historically velocity-dependent friction
forces were first studied  by Rayleigh and Helmholtz. Extensions of
these models were investigated in many papers on driven Brownian dynamics
\cite{Klim95}. 
A characteristic property of these friction functions 
is the existence  of a zero of friction for a finite velocity $v_0$ which
defines a kind of attractor in the velocity-space. 
The model we use here for active friction with an
attractor for the  velocities $v_0$
was introduced in \cite{ScEbTi98} and \cite{EbScTi99}. Detailed studies of this
so-called energy depot model model may be found in \cite{ErEbSS00,EbErDJ00}. 
The case
of Rayleigh-friction was analyzed in detail in \cite{MaEbVe00}. 
The depot-model of the friction function is
based on a concrete model of Brownian motion with energy supply, storage in a
tank and conversion into motion \cite{ScEbTi98}.  We assumed that the Brownian
particle itself is capable of taking up external energy storing some of this
additional energy into an internal energy depot, $e(t)$. 
Within the energy depot model the depot may
be may be altered by three different processes:
\begin{enumerate}
\item take-up of energy from the environment; where $q$ is the
  pump rate of energy
\item internal dissipation, which is assumed to be proportional to the
  internal energy. The rate of energy loss is denoted by $c$.
\item conversion of internal energy into motion, where $d$ is the rate
  of conversion of internal to kinetic degrees of freedom. The  
  depot energy is used to accelerate the motion.
\end{enumerate}

Our model of energy supply is motivated by investigations of active
biological motion, which relies on the supply of energy. 
The
supplied energy is in part dissipated by metabolic processes, but can be also
converted into kinetic energy. The energy depot model leads in an 
adiabatic approximation to the following friction function
\begin{equation}
  \gamma(v) = \gamma_0 - \frac{q}{1 + d v^2}
\end{equation}
For the energy depot model
of active friction exists an attracting velocity.
This is defined by the zero of the friction function 
$\gamma(v_0) = 0$ which
has the value 
\begin{equation}
v_0^2 = \frac{\alpha}{d}\,;\,\, \alpha=\frac{q}{\gamma_0}-1
\end{equation}
Here $\alpha$ plays the role of the bifurcation parameter. For
$\alpha < 0$ the friction is purely passive, i.e. in average no energy is
supplied. For $\alpha > 0$ the friction function has a negative part near to
$v = 0$ and a zero point which acts as an attracting set in the velocity
space.

The interaction between the masses $i$ and
$i+1$ is in the $1d$-case modelled by the Toda force
\begin{equation}
f_i = - V'(r_i) = \frac{\omega^2}{b}
(\exp \left[ -b r_i\right] -1)
\end{equation}
Here $b$ is the stiffness of the springs and $\omega$ is the linear
oscillation frequency around the equilibrium position.

In the harmonic limit the force is given by $f_i = \omega r_i$.
The spring energy is described by the Toda potential
\begin{equation}
V(r_i;\omega,b)=\frac{m\omega^2}{b^2}
(\exp \left[ -b r_i\right] -1+b r_i)
\end{equation}
We will use $m\omega^2 \sigma^2 $ as
the energy unit.\\ 
In general the calculation of the distribution functions of interacting
particles is not  a trivial task
since it is connected with the solution of a multi-dimensional 
Fokker-Planck equation.
A full solution is available only for the passive case $\gamma = \gamma_0 = const.$
Then the only attractor of the dynamics 
is the rest state of the particles and
the statistical properties in equilibrium are described by a 
canonical Toda ensemble under pressure $P$ and 
temperature $T$ \cite{BoOp81,ToSa83,Je91}. 
The stationary solution of the Fokker-Planck equation which corresponds 
to our Langevin equation reads
\begin{eqnarray} 
P_0 (p_n,r_n)= Z_1^{-1} \exp \left( -\frac{p_n ^2/2m+V_{eff}}{k_BT}\right) 
\nonumber \\ 
Z_1^{-1} = \frac{b X^{X+Y}}{\sqrt{2\pi mk_BT}\exp \left( X\right) 
\Gamma (X + Y)} 
\end{eqnarray} 
with the effective potential 
\begin{equation} 
V_{eff}(r_n)=V(r_n)+P r_n 
\end{equation}
Here $k_B$ denotes the Boltzmann constant. 
and $V(r)$ is the Toda potential defined above. 
Further we used the abbreviations 
\begin{equation}
X=\frac{m \omega^2}{b^2 k_BT}\,;\,\, Y=\frac P{b k_BT} 
\end{equation} 

\begin{figure}[b]
\begin{center}
\includegraphics[width=.6\textwidth]{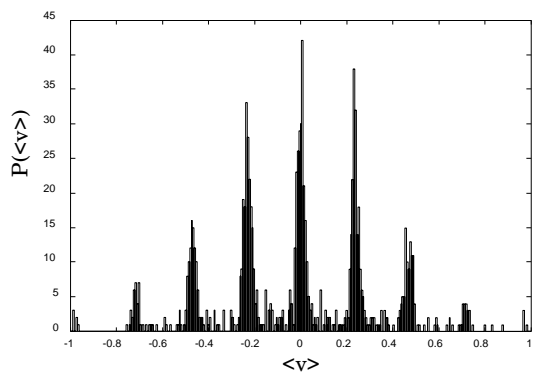}
\end{center}
\caption[]{Probability distribution for 8 active Brownian particles}
\label{Fig:Histo}
\end{figure}

The elementary excitations in passive Toda rings including the
noise spectrum and the structure factor were investigated
in detail in the work \cite{JeEb00,EbChJe00}.
The investigation of the dynamics of Toda rings with energy supply by 
active friction
is more difficult and only partial solutions are available
\cite{EbErDJ00,MaEbVe00}. We discuss here the special case when
the supply of energy is given by the depot model in the 
approximation of a 
velocity-dependent friction \cite{ScEbTi98,EbScTi99,ErEbSS00}.
Due to the driving slow particles are accelerated and fast particles are
damped. At definite conditions our active friction function has a zero
corresponding to stationary velocities $v_0$, where the friction disappears.
The deterministic trajectory of our system is then attracted by a
cylinder in the 4d-space given by
\begin{equation}
  v_1^2 + v_2^2 = v_0^2
\end{equation}
where $v_0$ is the value of the stationary velocity which is for the
depot-model given by $v_0^2=\frac{\alpha}{d}$.\\
In Fig. \ref{Fig:Histo}  we see the result of a histogram
calculated by J. Dunkel for $N = 8$ particles with 
Toda interactions \cite{EbErDJ00}. Here the result of 1000
runs with stochastic initial conditions is represented
as a function of the average velocity of the particles. The shape
of the histogram demonstrates the existence of 
$N + 1$ attractors of motion. We underline however that the
histogram shown in Fig.~1 is the result of finite time runs,
so it represents the shape of a finite-time distribution
function. In the limit of long runs the distribution may
change; we expect e.g. that at long times the statistical weights
are shifted from the center of the distribution to the wings. 

\section{Two-dimensional dynamics}

The effects demonstrated in the previous sections are not restricted 
to $1-d$ Toda lattices but persist at least qualitatively in more realistic
$2-d$ and $3-d$ models of  dense fluids consisting of solvent and solute
molecules with Morse- or  Lennard-Jones interactions. Superposition of solitons
corresponds to  multiple collisions in these systems. In higher dimensions a
weak localisation of  potential energy was observed also at the bindings of
the  bath molecules and was connected to a transition between different
lattice  configurations \cite{EbPoVa95,EbPo96,EbSaVa98}.\\
We introduce interactions described by the potential 
$U({\vec r}_1,...,{\vec r}_N)$; then the dynamics of Brownian particles 
is determined by the Langevin equation:
\begin{equation}
  \label{langev-or}
  \dot{{vec r}_i}={\vec v}_i\,;\,\, 
  m\dot{{\vec v}_i}=-\gamma(v_i){\vec v}_i -
\nabla U({\vec r}_1,...,{\vec r}_N) + {\vec {\cal F}}(t) 
\end{equation}
where 
$\cal{F(t)}$ is a stochastic force with strength $D$ and a
$\delta$-correlated time dependence as defined above.\\
We will discuss now the motion of active particles in a two-dimensional space,
$\vec r=\{x_{1},x_{2}\}$. The case of constant external forces was alredy
treated by Schienbein and Gruler \cite{ScGr93}. 
Symmetric parabolic external forces were studied in 
\cite{EbScTi99,ErEbSS00,EbErDJ00}
and the non-symmetric case is beeing investigated in {\cite{AnErEb00}.

Here we will study 2d-systems of $N \geq 2$ particles 
(see Fig. \ref{swarm1})
\begin{figure}[b]
\begin{center}
\includegraphics[width=.5\textwidth]{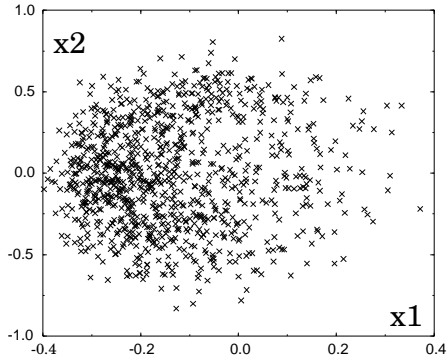}
\end{center}
\caption[]{Snapshot of 1000 Brownian particles rotating in a parabolic
self-consistent field} 
\label{swarm1}
\end{figure}

We will begin with the study of a 2 particle problem \cite{AnErEb00}. 
Let us imagine two Brownian particles 
which are pairwise bound by a pair
potential $U({\vec r}_1 - {\vec r}_2)$. The two molecules will form
dumb-bell-like configurations.  Then the motion
consists of two independent parts: The free motion of the center of mass 
having the coordinates $X_1 = 0.5 (x_{11} + x_{21})$ and
$X_2 = 0.5 (x_{12} + x_{22})$. The relative motion under
the influence of the potential is described by the
coordinates $\tilde x_1 = (x_{11}-x_{12})$ and
$\tilde x_2 = (x_{12} -  x_{21})$.  
The motion of the center of mass $M$ is described by the equations

\begin{eqnarray}
    \dot{X_1}  = V_1 & \qquad &
    M \dot{V_1} = - \gamma\left(V_1,V_2\right) V_{1} + {\cal F}_1(t) \nonumber\\ 
    \dot{X}_2  = V_2 & \qquad &
    M \dot{V}_2 = - \gamma\left(V_1,V_2\right) V_2  +{\cal F}_2(t)
\label{2d-center}
\end{eqnarray}

The stationary solutions of the corresponding Fokker-Planck 
equation reads \cite{ErEbSS00}

\begin{equation}
P_0(\vec V) = C \left(1 + d  V^2\right)^{(q/2D)} 
\exp\left[ -\frac{\gamma_0} {2D} \, V^2\right]
\end{equation}

This corresponds to the driven motion of a free particle located in the 
center of mass.\\
The relative motion is described by the equations

\begin{eqnarray}
    \dot{\tilde x}_1  = \tilde v_1 & \qquad &
    \mu \dot{\tilde v}_1 = - \gamma\left(\tilde v_1,\tilde v_2\right) 
\tilde v_1 -  \partial_1 U  + {\tilde {\cal F}}_1 (t)
    \nonumber\\ 
    \dot{\tilde x}_2  = \tilde v_2 & \qquad &
    \mu \dot{\tilde v}_2 = - \gamma\left(\tilde v_1,\tilde v_2\right) 
\tilde v_2 -      \partial_2 U + {\tilde {\cal F}}_2 (t)
\label{2d-relat}
\end{eqnarray}

with $\mu = m/2$.
For simplification we specify
now the potential $U(\vec r)$ as a symmetric parabolic potential 
\cite{EbErDJ00}:

\begin{equation}
U(x_1,x_2) = \frac{1}{2} a \,(x_1^2 + x_2^2)
\label{parab}
\end{equation}

First, we restrict the discussion to a deterministic relative motion, which
is described by four coupled first-order differential equations.
The relative motion of 2 particles corresponds to the absolute motion of 1 particle
in an external field. Therefore we will omit now for simplicity the tilde
denoting the relative character of the motion and denoting the mass
again by $m$ instead of $\mu$ :

\begin{eqnarray}
    \dot{x}_1  = v_1  \qquad 
    m\dot{v}_1 = - \gamma\left(v_1,v_2\right) v_{1} - ax_{1} \nonumber\\ 
    \dot{x}_2  = v_2  \qquad 
    m\dot{v}_2 = - \gamma\left(v_1,v_2\right) v_2 - a x_{2}
\label{2d-det}
\end{eqnarray}

For the one-dimensional case it is well known that this
system pocesses a limit cycle corresponding to sustained oscillations
\cite{Klim95}.
For the $2-d$ case we have shown in \cite{EbScTi99} that a limit cycle in the
$4-d$ space is developed, which corresponds to left/right rotations
with the frequency $\omega_0$. The
projection of this periodic motion to the $\{v_1,v_2\}$ plane 
is the circle

\begin{equation}
  v_1^2 + v_2^2 = v_0^2 = {\rm const.}
\end{equation}

The projection to the $\{x_1,x_2\}$ plane also corresponds to a circle
\begin{equation}
  x_1^2 + x_2^2 =  r_0^2 =  \frac{v_0^2}{\omega_0^2}
\end{equation} 

The energy for motions on the limit cycle is
\begin{eqnarray}
  E_0 &=& \frac{m}{2}(v_1^2 + v_2^2) +\frac{a}{2} (x_1^2 + x_2^2)\\
  &=& \frac{m}{2} v_0^2 +\frac{a}{2} r_0^2
\end{eqnarray}

We have shown in \cite{EbScTi99}, that any initial value of the energy
converges (at least in the limit of strong driving) to
\begin{equation}
  H\longrightarrow E_0 = m v_0^2
\end{equation}

This corresponds to an equal distribution between kinetic and potential
energy. As for the harmonic oscillator in $1-d$, both parts contribute the same
amount to the full energy. This result was obtained in \cite{EbScTi99} based
on the assumption that the energy is a slow (adiabatic) variable which allows
a phase average with respect to the phases of the rotation. In explicite form
we may represent the motion on the limit cycle in the $4-d$ space by the 4
equations \cite{ErEbSS00}

\begin{equation}
    x_1 = r_0\, \sin(\omega_0 t) \,;\,\,
    v_1 = - r_0\, \omega_0\, \cos(\omega_0 t)
\end{equation}    
\begin{equation}
    x_2 = r_0\, \cos(\omega_0 t)\,;\,\,
    v_2 = - r_0\, \omega_0\, \sin(\omega t)
\end{equation}

The frequency is given by the time the particle need for one period moving on
the circle with radius $r_0$ with constant speed $v_0$. This leads to the
relation

\begin{equation}
  \omega_0 = \frac{r_0}{v_0} = \left(\frac{m}{a}\right)^{1/2} = \omega
\end{equation}

This means, the particle oscillates (at least in our
approximation) with the frequency given by the linear oscillator frequency
$\omega$.

The trajectory on the limit cycle defined by the above 4 equations is like a
hula hoop in the $4-d$ space. The projections to the $x_1 - x_2$-space as well
as the projections to  the $v_1-v_2$-space are
circles.
The projections to the subspaces 
$x_1$-$v_2$ and $x_2$-$v_1$ are like a rod. In the $4-d$ space the attractor
has therefore the form of a hula hoop.
A second limit cycle is obtained by reversal of the velocity.
This second limit cycle forms also a hula hoop which is different from the 
first one, however both l.c. have the same projections to the $\{x_1,x_2\}$
and to the $\{v_1,v_2\}$ plane. The motion in the $\{x_1,x_2\}$
plane has the opposite sense of rotation in comparision with the first limit
cycle.  Therefore both limit cycles correspond to opposite angular momenta.
$L_3 = +\mu r_0 v_0$ and $L_3 = -\mu r_0 v_0$.
Applying similar arguments to the stochastic problem we find that the
two hoop-rings are converted into a distribution looking like two embracing
hoops with finite size, which for strong noise converts into two embracing
tyres in the $4-d$ space.
In order to get the explicite form of the
distribution we may introduce the amplitude--phase representation
\cite{ErEbSS00} 

\begin{equation}    
    x_1 = \rho sin(\omega_0 t + \phi) \,;\,\,
    v_1 = -\rho \omega_0 cos(\omega_0 t + \phi)
\end{equation}
\begin{equation}
    x_2 = \rho cos(\omega_0 t + \phi)\,;\,\,
    v_2 = - \rho \omega_0 sin(\omega_0 t + \phi)
\end{equation}

where the radius $\rho$ is now a slow and the phase $\phi$ is a fast
stochastic variable.  By using the standard procedure of averaging with
respect to the fast phases we get the distribution of the radii
\cite{ErEbSS00}.

\begin{equation}
\label{solFP}
P_{0}(\rho) = C \left(1 + d {\rho \omega_0}^2 \right)^{(q/2D)} 
\exp\left[ -\frac{\gamma_0}{2D} {\rho \omega_0}^2 \right]
\end{equation}

The probability crater is located above the two deterministic
limit cycles on the sphere $r_0^2 = v_0^2 /\omega_0^2$. Strictly speaking not
the whole sherical set is filled with probability but only two circle-shaped
subsets on it, which correspond to a narrow region around the limit sets. The
full stationary probability has the form of two hula hoop distributions in the
4d space. This was confirmed by simulations \cite{ErEbSS00}.

The projections of the distribution to the $\{x_1,x_2\}$ plane and to the
$\{v_1,v_2\}$ plane are smoothed $2d$-rings. The distributions
intersect perpendicular the $\{x_1,v_2\}$ plane and the $\{x_2,v_1\}$ plane.
Due to the noise the Brownian particles may switch between the two limit cycles,
this means inversion of the angular momentum (direction of rotation)
\cite{ErEbSS00,AnErEb00}.\\

Summarizing our findings for a 2-particle system forming a dumb-bell system
we may state:
The center of mass of the dumb-bell will make a driven
Brownian motion corresponding to a free motion of the center of mass. 
In addition the dumb-bell is driven to rotate around
the center of mass. What we observe then is a system of rotating Brownian
molecules. The internal degrees
of freedom are excited and we observe driven rotations.

In this way we
have shown that the mechanisms described here may be used also to excite the
internal degrees of freedom of Brownian molecules.\\

An extension of the theory of pairs leads to a theory of the motion of 
clusters of active molecules \cite{ViCz95,DeVi95,AnErEb00,EbScTi00}.
In Fig. \ref{swarm2} we see a snapshot of simulations of the stochastic
dynamics of a cluster of 1000 active interacting Brownian particles.
\begin{figure}[b]
\begin{center}
\includegraphics[width=.5\textwidth]{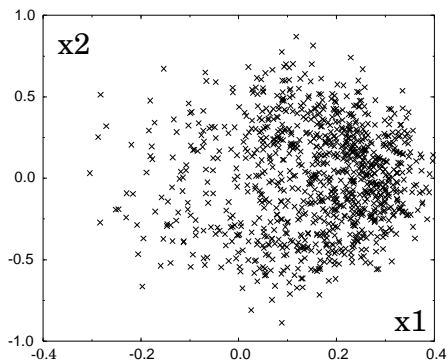}
\end{center}
\caption[]{Snapshot of the same swarm after one half period of rotation} 
\label{swarm2}
\end{figure}

In order to simplify the simulations we assumed that the interaction of
the molecules in the cluster is given by a van der Waals - like interaction
with a relatively long range range.  For example we may use the interaction
model proposed by Morse
\begin{equation}
\phi(r) = A \left[\exp(-a r) -1 \right]^2 - A 
\end{equation}
Due to the attracting tail the molecules form clusters. The individual 
molecules move then in the collectice (self-consistent) 
field of the other molecules which might be represented 
by a mean field approximation 
\begin{equation}
V (\tilde{\vec r}) = \int d \vec r' \phi(\tilde{\vec r} - \vec r') \rho(\vec r')
\end{equation}
where $\tilde{\vec r} = (\tilde x_1, \tilde x_2)$ is the radius vector counted from the center of 
mass and $\rho(\vec r')$ is a mean density in the cluster. 
Approximating $V$ by quadratic terms only we get
\begin{equation}
V (\tilde x_1, \tilde x_2 ) = V_0 + \frac{1}{2} 
\left(a_1 \tilde x_1^2 +  a_2 \tilde x_2^2 \right) + ...
\end{equation} 
\begin{figure}[b]
\begin{center}
\vspace{3cm}
\includegraphics[width=.5\textwidth,angle=-90]{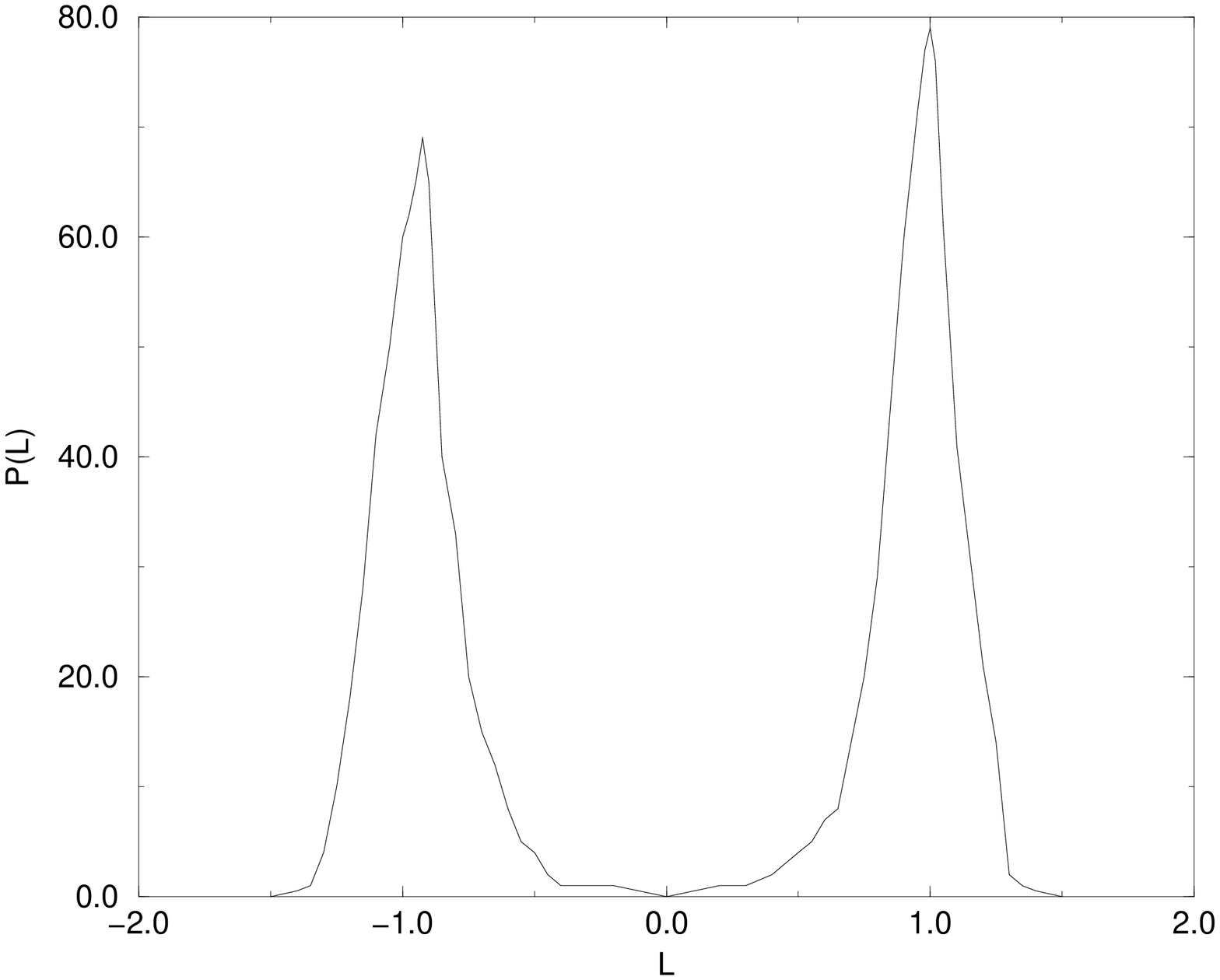}
\end{center}
\caption[]{Probability distribution of the angular momentum for active
Brownian particles with $v_0^2 = r_0^2 = 1$}
\label{angmo}
\end{figure}

In this way we arrive again at the harmonic problem we have studied above. 
We may conclued that the individual molecules in the cluster move at 
least in certain approximation in a parabolic potential.
From this follows that they will make rotations in the field.
We have performed
simulations with 1000 particles moving in a self-consistent potential of
parabolic shape. The driving  function has a zero at $v_0^2 =1$. In Figs.
\ref{swarm1} and \ref{swarm2} one sees two snapshots of the moving cluster
formed by the molecules.

Since the indidual particles move in an effective parabolic potential 
angular momenta are generated, the swarm starts to rotate.
Similar as in the case of the dumb-bells the clusters are driven to make
spontaneous rotations. Finally a stationary state will be reached which 
corresponds to a rotating cluster with nearly constant angular momentum 
(see Fig. \ref{angmo}). Under the influence of noise the cluster may 
switch to the opposite angular momentum, i.e. to the opposite sense of
rotation, the system occurs to be bistable. 

In Fig. 4 is shown that in noisy systems of active Brownian particles the two 
values $L_3 = -mr_0 v_0, +mr_0 v_0$ have the maximal probability.
Strong coupling of the particles leads to synchronization of the 
angular momenta, for weak coupling the cluster may be decomposed
into groups with different angular momentum \cite{AnErEb00,EbScTi00}. 
The rotating swarms simulated in our numerical experiments remind very much
the dynamics of swarms studied in papers of Viscek and collaborators
\cite{ViCz95,DeVi95} and in other recent work \cite{AnErEb00,EbScTi00}.

\section{Conclusions}

We studied here the active Brownian dynamics of a finite number of particles
including interactions 
modelled by Toda or Morse potentials and velocity-dependent friction.
Beside analytical investigations we have made a numerical study
of relatively small systems. 
For this we have been investigating $1d$-model systems with
nonlinear Toda interactions and $2d$-models with parabolic confinement. 
At low enough temperatures we observe only harmonic excitations and the 
models reduce to systems of coupled harmonic oscillators.
However at higher temperatures the dynamics is completely changed
and nonlinear excitations of soliton-type come into play. 
At high temperatures the behaviour of $1d$ interacting systems is 
dominated by soliton-type excitations. Including active friction we 
observe in the $1d$-case $N+1$ attractors of
nonlinear excitations. For interacting Brownian particles in $2d$-case
angular momenta are generated (see Fig. 4) and 
we find rotations of the particles around the center of attraction
and collective rotational excitations of swarms.

Acknowledgements:
The author is grateful to A. Chetverikov (Saratov), J. Dunkel and U. Erdmann
(Berlin), M. Jenssen (Eberswalde),  Yu. L. Klimontovich (Moscow),
F.~ Schweitzer~ (Birlinghoven), B.~ Tilch (Stuttgart), 
V. Makarov and M. Velarde (Madrid) for collaboration and support of this
work. The FORTRAN program used for generating Figs. 2-4 was written 
by A. Neiman (St. Louis).

\end{document}